# Superconductivity in multiple phases of compressed $GeSb_2Te_4$

E. Greenberg [1†], B. Hen [1†], Samar Layek [1†], I. Pozin [2], R. Friedman [3], V. Shelukhin [1], Y. Rosenberg [4], M. Karpovski [1], M. P. Pasternak[1], E. Sterer [3], Y. Dagan [1], G. Kh. Rozenberg [1] and A. Palevski [1*]

[1] *Raymond and Beverly Sackler School of Physics and Astronomy, Tel-Aviv University, Tel Aviv 69978, Israel*

[2] *Department of Material Science and Engineering, Tel-Aviv University, Tel Aviv 69978, Israel*

[3] *Physics Department, Nuclear Research Center Negev, P.O. Box 9001, Beer-Sheva 84190,Israel*

[4] *Wolfson Applied Material Research Center, Tel-Aviv University, Tel Aviv 69978, Israel*

Here we report the discovery of superconductivity in multiple phases of the compressed $GeSb_2Te_4$ (GST) phase change memory alloy, which has attracted considerable attention for the last decade due to its unusual physical properties with many potential applications. Superconductivity is observed through electrical transport measurements, both for the amorphous (a-GST) and for the crystalline (c-GST) phases. The superconducting critical temperature, TC, continuously increases with the applied pressure reaching a maximum $T_c$ =6K at P=20 GPa for a-GST, whereas the critical temperature of the cubic phase reaches a maximum $T_c$ =8 K at 30 GPa. This new material system, exhibiting a superconductor-insulator quantum phase transition (SIT) has an advantage over disordered metals since it has a continuous control of the crystal structure and the electronic properties using pressure as an external stimulus, which was lacking in SIT studies until today.

In conventional metals and alloys coulomb interactions are screened by the conducting electrons but an attraction between electrons due to the electron-phonon interaction results in Superconductivity, which appears at a critical temperature $T_c$ determined by the Debye frequency and the electron-phonon coupling constant. A system where the crystal structure can be tuned and the material can undergo a transition from a metallic to an insulating state by applying an external stimulus can therefore become a laboratory for studying fundamental questions about superconductivity.

$GeSb_2Te_4$ (GST) is a phase change material whose unusual physical properties [1, 2, 3, 4, 5, 6] promise many potential applications in the electronics industry [7, 8, 9, 10]**.** GST can undergo a reversible change from an amorphous phase (*a*-GST) to a crystalline (*c*-GST) one at elevated temperatures [1, 2, 3, 4], but also by elevated pressure at ambient temperature [11]. This transition is accompanied by a significant change of over 4 decades in resistance. The observed resistance changes, as well as reversible *a*-GST to *c*-GST phase transition, have been explained using numerical simulations [11, 12, 13]. The corresponding metal-to-insulator transition (MIT) in GST is often explained as a disorder-induced Anderson localization [1].

Superconductivity near a metal-to-insulator transition has been reported in disordered metals [14] and has been studied intensively. Nevertheless, a continuous control of the crystal structure and the electronic properties using external stimuli was lacking until today. Here, using pressure as an external stimulus, we show that GST becomes superconducting at low temperature. We study the superconducting transition in the temperature-pressure phase diagram and demonstrate that superconductivity is observed both for the amorphous (*a*-GST) and for the crystalline (*c*-GST) phases.

In our transport studies of GST material under the pressure in a wide temperature range, we used as grown amorphous (*a*-GST) and as prepared crystalline (*c*-GST) powders; pressure was generated using diamond anvil cells (DACs); electrical transport measurements were performed using a physical property measurement system (for details see Supplementary methods). The lattice of *c*-GST is hexagonal at ambient temperature and atmospheric pressure, and therefore the crystalline samples are denoted as *h*-GST in their initial conditions. The X-ray diffraction analysis of our amorphous and crystalline samples, shown in the supplementary materials, supplementary Fig. 1, confirms the amorphous structure of as-prepared *a*-GST and the hexagonal structure of *h*-GST. Upon application of the pressure at ambient temperature we observe, in agreement with Ref. [15], a gradual structural transformation from the amorphous phase into a high density amorphous phase at around 10 GPa followed by the crystallization into a *bcc* lattice at pressures above 20 GPa. We also observe, all phases of crystallographic transitions reported in Ref. [16] for the initially crystalline *h*-GST (see supplementary Fig. 1)

Two cells of *a*-GST were prepared (samples 1 and 2), one made of nonmagnetic materials to enable measurements at various magnetic fields (sample 1). Two cells with *h*-GST sample were also measured as a function of temperature for various pressures. In Fig. 1(a), we show the typical contact configuration for measuring resistance at various pressures as a function of temperature.

At ambient temperature, the *a*-GST samples show a dramatic, over five orders of magnitude, drop in the resistance, when it is compressed applying quasi-hydrostatic pressure, followed by saturation above 9 GPa as shown in Fig. 1(b). The behavior of *h*-GST also saturates above 10 GPa as seen Fig. 1(c), however, the overall change in resistance is much smaller.

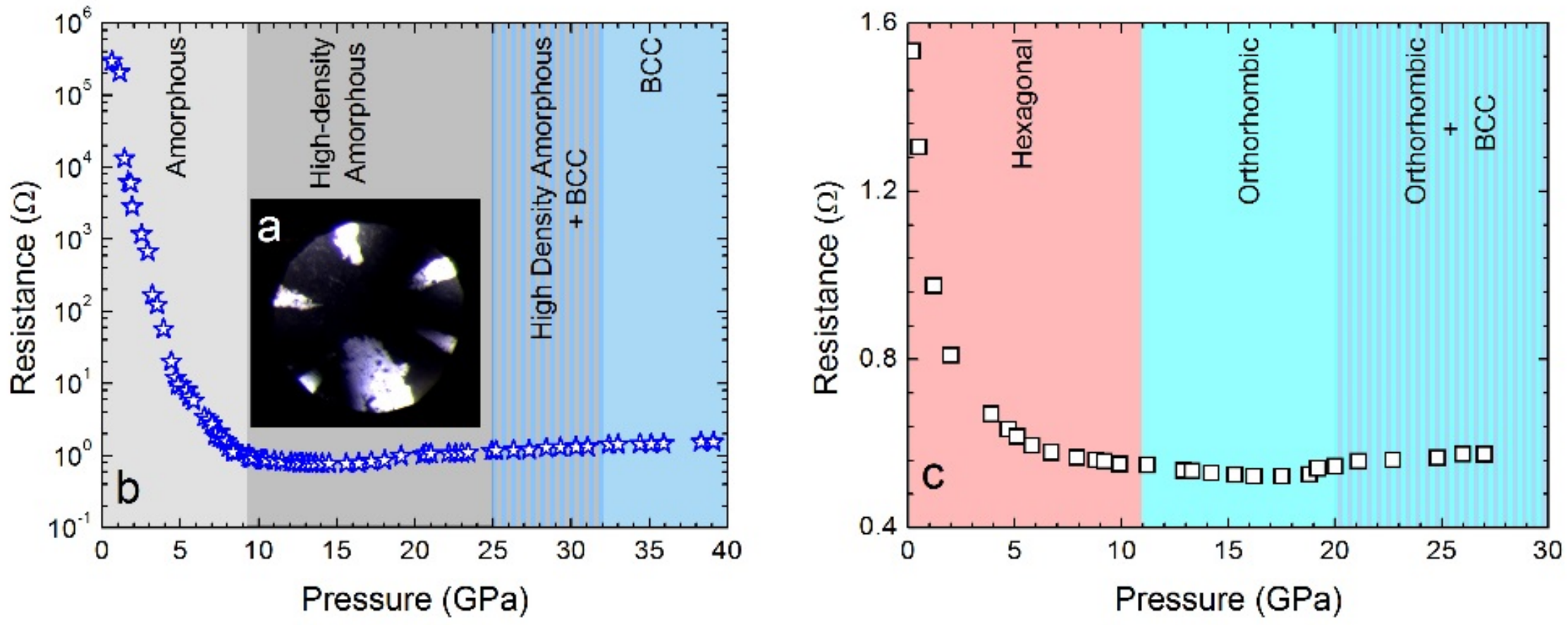

FIG. 1. Room temperature resistance under pressure. (a) Six platinum triangles under pressure in one of the measurement set-ups. (b) Resistance as a function of pressure of *a*-GST. (c) Pressure dependence of resistance of *c*-GST. Shaded regions in different colors (in both (b) and (c)) are according to previously published XRD data [15, 16] and our resistance measurements.

The dramatic drop in resistance for *a*-GST sample is corroborated with the corresponding change of the mechanism of conductivity. For low pressures, below 5 GPa the variation of the resistance versus temperature indicates that the *a*-GST is an Anderson insulator with a typical transition from a simply activated to Mott variable range hopping mechanism of the electronic transport. (For more information see supplementary material, supplementary Fig. 2). At pressures exceeding 9.5 GPa, when the resistivity drops below approximately 1 mΩ-cm, the value reported previously [1] as a precursor of the metal insulator transition (MIT), the *a*-GST becomes a superconductor.

In Fig. 2 we show the resistance versus temperature for both samples. The plots in Fig. 2 show that for pressures below P=21 GPa, there is a monotonous increase of the superconductor transition temperature with pressure, accompanied with a decrease of the normal state resistance. The behavior changes drastically when the pressure is increased above 21 GPa. One can clearly see that the curves have two distinct transitions, signifying the appearance of an additional phase with a higher transition temperature. This double transition is observed in both samples. As demonstrated by our XRD data in the supplementary material, and as was also reported previously by others [15], at this range of pressure, a crystallographic phase transition occurs forming the bcc ordered phase, bcc-GST. We can, therefore, interpret the double transition as coexistence of a-GST with the bcc-GST, both being superconductors with a higher value of Tc for the crystalline phase.

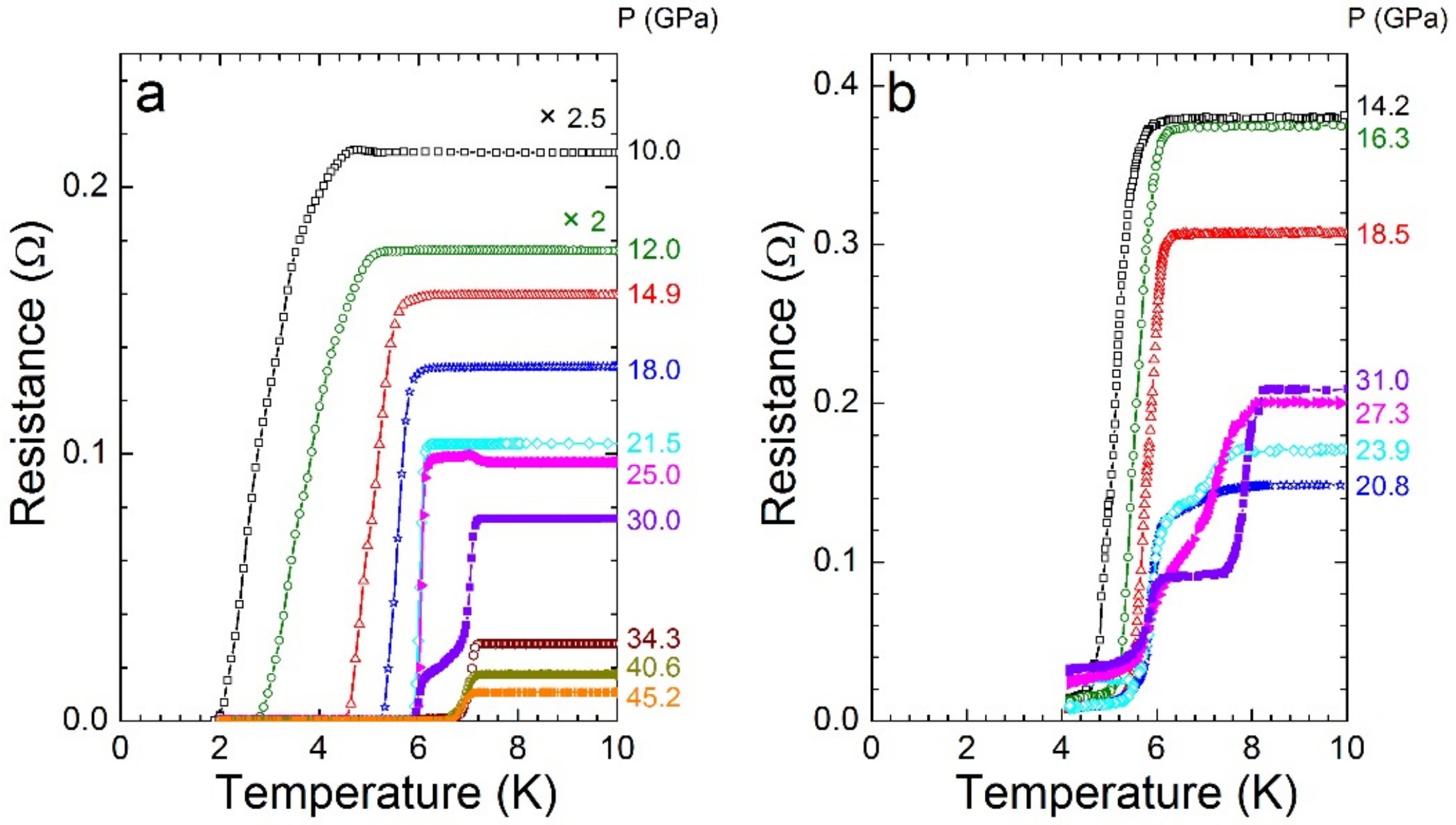

FIG. 2. Superconducting transitions of *a*-GST. (a) Temperature dependence of resistance at different pressures using four probe geometry (Sample 1). (b) (Sample 2) using quasi four probe geometry. Two distinct superconducting transitions are observed.

In Fig. 3(a), we show magnetoresistance measurements at 2 K for sample 1 at different pressures. From these curves we can extract the upper critical field $H_{C2}$ defined as the field for which the resistance attains half of the normal state resistance value. A closer inspection of the curves shown in Fig. 3(b) reveals a double transition versus magnetic field. These double transitions in the magnetic field appear at somewhat higher pressure than corresponding double transitions observed in $R(T)$ curves. For these pressures, exceeding 40 GPa, only single *bcc* phase is expected to remain. This inconsistency is the subject for further investigation. The values of the corresponding upper critical field are plotted in Fig. 3(c).

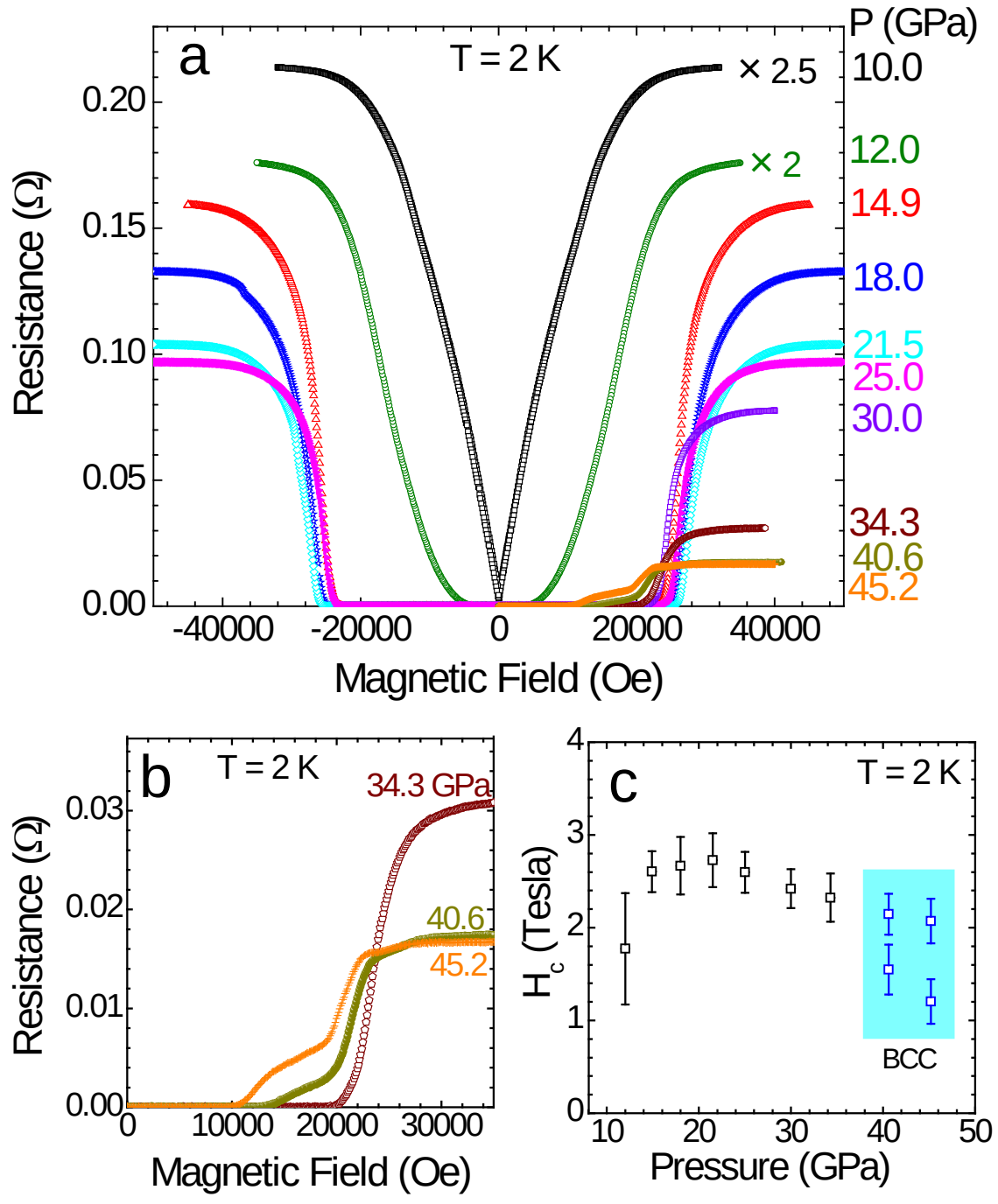

FIG. 3. Pressure dependence of the critical field of *a*-GST (sample 1). (a) Resistance as a function of magnetic field at different pressures. (b) Zoomed portion of the transitions from (a) for 34.3, 40.6 and 45.2 GPa showing two transitions in the last two pressures. (c) Variation of the critical field as a function of pressure.

Our definition of $H_{C2}$ is not appropriate for the pressure range where we observe a double transition (coexistence regime). In the latter case we estimated $H_{C2}$ as the mid-point value of each transition for each phase. Using the typical value of $H_{C2}$=2.8 T within the Ginzburg Landau and BCS formalism [17] we estimate the following microscopic parameters as follows. The Ginzburg Landau coherence length, $\xi_{GL}(T) \approx 30\ nm$ at $T \approx 2\ K$. Assuming that our samples are in a clean limit, $\ell \geq \xi_0$ , which is a reasonable assumption for the metallic GST with the resistivity $\rho \leq 10\ \mu\Omega\ cm$ we can estimate the Pippard coherence length at low temperatures to be $\xi_0 \approx 40\ nm$. If we assume that superconducting gap follows BCS theory, $\Delta = 3.5\ k_B T_c$, then we can estimate the Fermi velocity using the relation: $\xi_0 = \frac{\hbar v_f}{\pi \Delta}$, taking $T_C$=7 K for *bcc*-GST, we find, $v_f \approx 3 \cdot 10^5 m/s$ . Additional current-voltage measurements can be found in the supplementary information, supplementary Fig. 3.

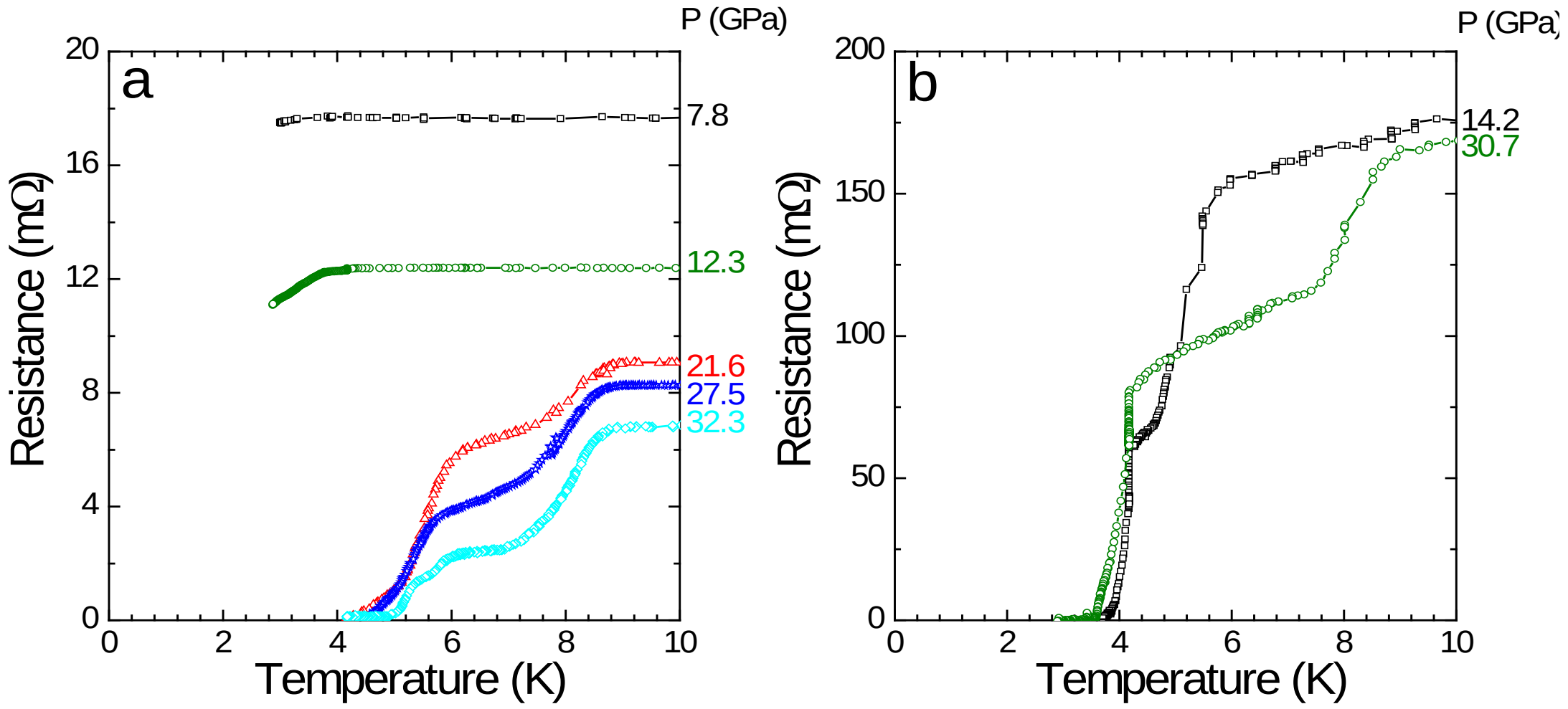

FIG. 4. Superconducting transitions of *c*-GST. (a) Temperature dependence of resistance at different pressure. (b) Same measurements using two wires after subtracting residual resistance values (1.329 Ω and 1.564 Ω for 14.2 and 30.7 GPa, respectively).

We also performed low temperature studies on two samples prepared in the hexagonal phase, which were measured in different DACs. As we already mentioned, the behavior of *h*-GST under pressure is quite different: First, the resistivity changes by less than an order of magnitude at 300 K when compressed, see Fig. 1(b) and second, it exhibits normal metallic behavior at pressures below 10 GPa. The first crystalline sample was compressed at room temperature to 28 GPa, decompressed and recompressed again. The variation of the room temperature resistance upon this cycle is depicted in supplementary Fig. 4. Since our as prepared a-GST were superconducting when transformed into the bcc phase, one could expect that as prepared h-GST sample will also exhibit superconductivity upon compression following decompression from 28 GPa, where bcc-GST is formed according to supplementary Fig. 1 and Ref. [16]. We indeed observe the appearance of the superconductivity upon recompression at pressures above 14 GPa as shown in Fig. 4 (b). There are two separate transitions in the R(T) curve, which are quite distinct. We attribute the highest Tc to the bcc phase, since it is very close to the value of TC observed in as prepared a-GST samples undergoing the crystallographic transition to bcc phase, whereas the lower transition is attributed to the superconductivity in the orthorhombic phase (see supplementary Fig. 1 and Ref. [16]).

The measurements of the second h-GST sample were performed only in the compression mode. These measurements reveal similar behavior to the first sample behavior, i.e. they do not exhibit superconductivity at temperatures around 4 K for pressures below 14 GPa (Fig. 4(a)). At higher pressures, where h-GST transforms into different crystallographic phases (supplementary Fig. 1), we observe superconductivity at the same temperatures as the first sample upon recompression, see Fig. 4(b).

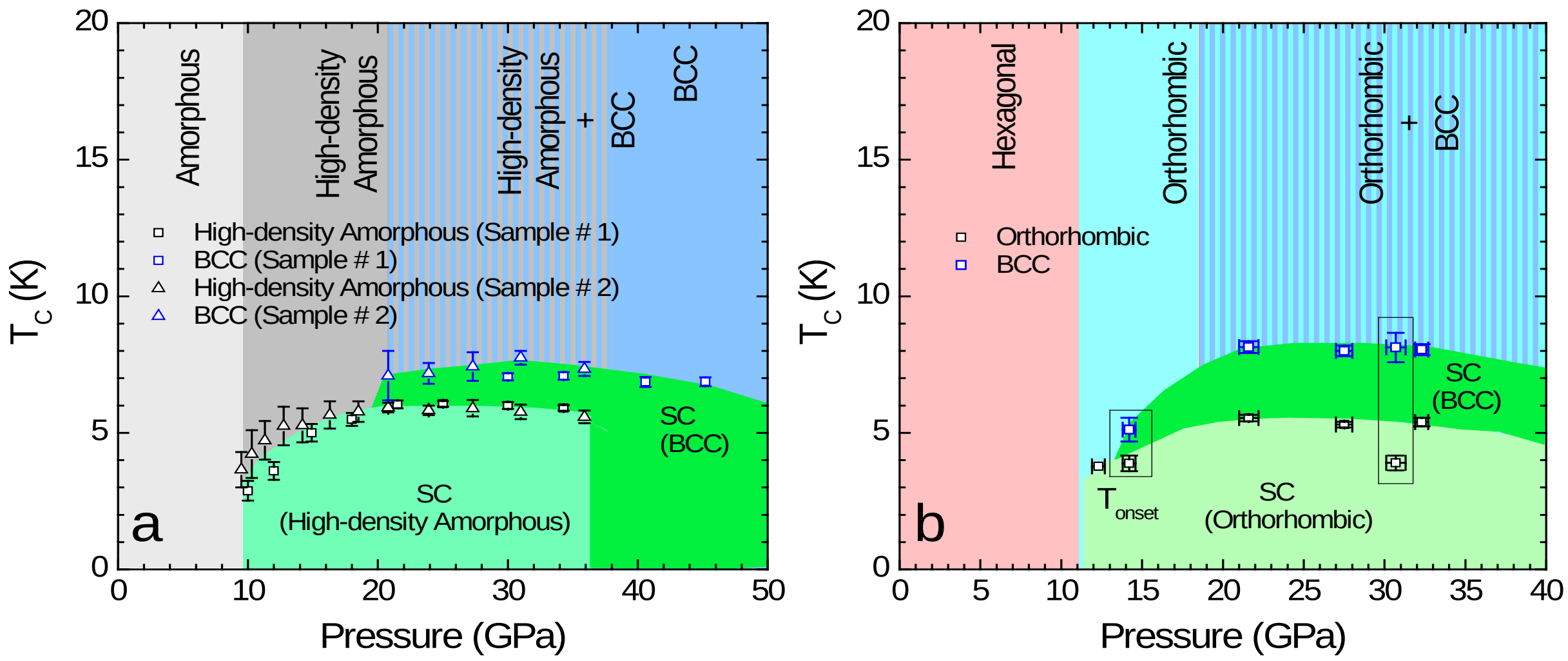

FIG. 5. Superconducting phase diagram of (a) a-GST and (b) c-GST. Shaded regions in different colors are according to our superconductivity measurements. Since the superconducting transition is not completed at 12.3 GPa which is the first point in (b), only onset temperature is given. The points inside rectangles in (b) are measured during re-compression cycle.

We use a similar definition for the critical temperature, TC as we used for defining HC2, namely as the temperature for which the value of the resistance is one half of its value at the normal state, prior to the transition. We can get a T-P phase diagram, which summarizes our findings for both as prepared a-GST and h-GST samples, by plotting the variation of the TC versus P in Fig. 5 for both measurement sets. At the range presented in our phase diagram it appears that the low density amorphous and the hexagonal phases are not exhibiting superconductivity. Much lower temperatures than those available in our current experimental set up are needed to complete the phase diagram.

In conclusion, we have demonstrated superconductivity in both amorphous and a few crystalline phases of GeSb2Te4 compound with the maximum transition temperature of Tc=8 K at a pressure of 30 GPa. The different superconducting transitions are well correlated with the observed structural transitions. We believe that in addition to being a phase change material, GST becomes a laboratory for studying superconductor-to-insulator transitions with the pressure being an external tuning parameter. It also provides a unique opportunity to correlate the quantum phase transition with crystallographic transformations.

## Acknowledgements

The authors would like to thank M. Shulman for his help with the VTI system and A. Rabinowicz, A. Ron and E. Maniv for their assistance. We are grateful to Volodymyr Svitlyk and Andrew Cairns at the ESRF for providing assistance in using beamline ID27. This research was partially funded by PAZI foundation under grant number 268/15. Support from the Israeli Science Foundation under grant numbers 569/13 and 1189/14, and from the Israel Ministry of Science, Technology and Space under contract number 3-11875 are acknowledged.

†E.G., B.H. and S.L. Equally contributed to this work.

---

* apalev@post.tau.ac.il

# Superconductivity in multiple phases of compressed $GeSb_2Te_4$

E. Greenberg [1†] , B. Hen [1†], Samar Layek [1†], I. Pozin [2], R. Friedman [3] ,V. Shelukhin [1], Y. Rosenberg [4], M. Karpovski [1], M. P. Pasternak[1], E. Sterer [3], Y. Dagan [1], G. Rozenberg [1] and A. Palevski [1*]

[1] *Raymond and Beverly Sackler School of Physics and Astronomy, Tel-Aviv University, Tel Aviv 69978, Israel*
[2] *Department of Material Science and Engineering, Tel-Aviv University, Tel Aviv 69978, Israel*
[3] *Physics Department, Nuclear Research Center Negev, P.O. Box 9001, Beer-Sheva 84190,Israel*
[4] *Wolfson Applied Material Research Center, Tel-Aviv University, Tel Aviv 69978, Israel*

## *Supporting Materials*

## Supplementary Figures

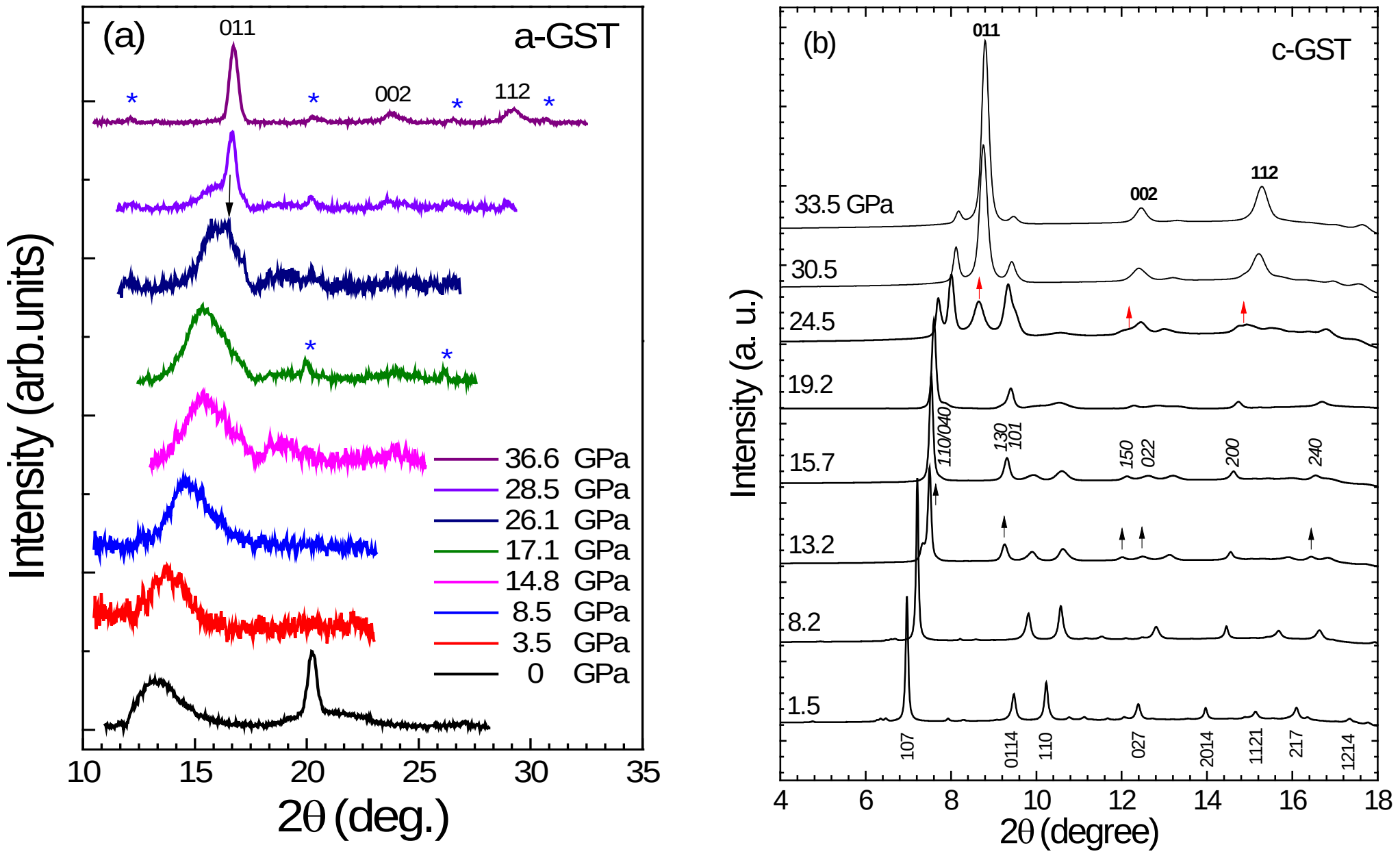

**Supplementary figure S1. X-ray powder diffraction spectra of GST at *T*=298 K at various pressures obtained upon compression. a,** *a*-GST. Downwards "↓" facing arrow demonstrates the appearance of the *bcc* phase at 26.1 GPa; Miller indices correspond to the diffraction peaks of this phase. The Bragg peak at ambient pressure at 20.3° corresponds to the scattering from the gasket; the asterisks correspond to the scattering from the ruby. **b,** *c*-GST. Upwards "↑" facing black and red arrows represent the appearance of the orthorhombic and *bcc* phases at 13.2 and 24.5 GPa, respectively; Miller indices in normal, italic and bold correspond to the diffraction peaks of the hexagonal, orthorhombic and bcc phases, respectively.

The diffraction patterns show that the *a*-GST sample remained amorphous to ~26 GPa. Up to about 14 GPa the amorphous halo shifts considerably, indicating a densification of the material under pressure. With further pressure increase the position of the halo barely changes suggesting transition from low density amorphous to high density amorphous phase (see [15]). A further increase in pressure results in the appearance of a small fraction of a crystalline phase while the sample remains largely amorphous. At about 36 GPa, the amorphous to crystalline phase transition is completed. The high pressure crystalline phase displays *bcc* symmetry and similarly to [15] can be indexed with the space group *Im-3m* with a unit cell dimension of a = 3.5001(11) Å at 29.2 GPa.

For *c*-GST sample, similarly to [16] the hexagonal phase remains stable up to 8.2 GPa. At 13.2 GPa, the orthorhombic phase appears coexisting with h-GST up to 15.7 GPa. At 24.5 GPa the onset of the bcc phase is observed, with abundance increasing gradually with pressure.

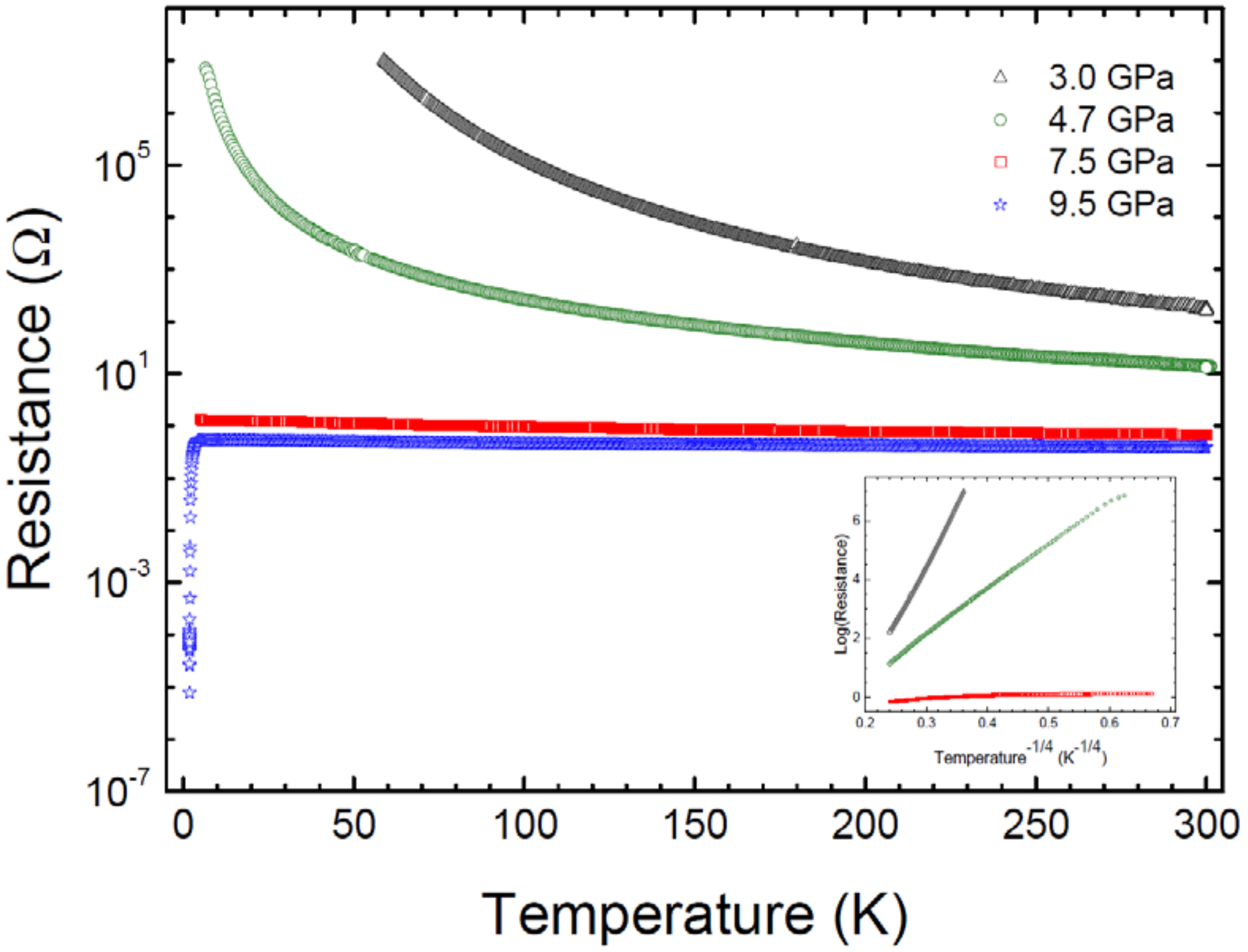

**Supplementary figure S2. Temperature dependence of a-GST samples demonstrating superconductor insulator transition (SIT).** The Mott variable range hopping is clearly observed at pressures between 3.0 and 7.5 GPa as seen from the inset.

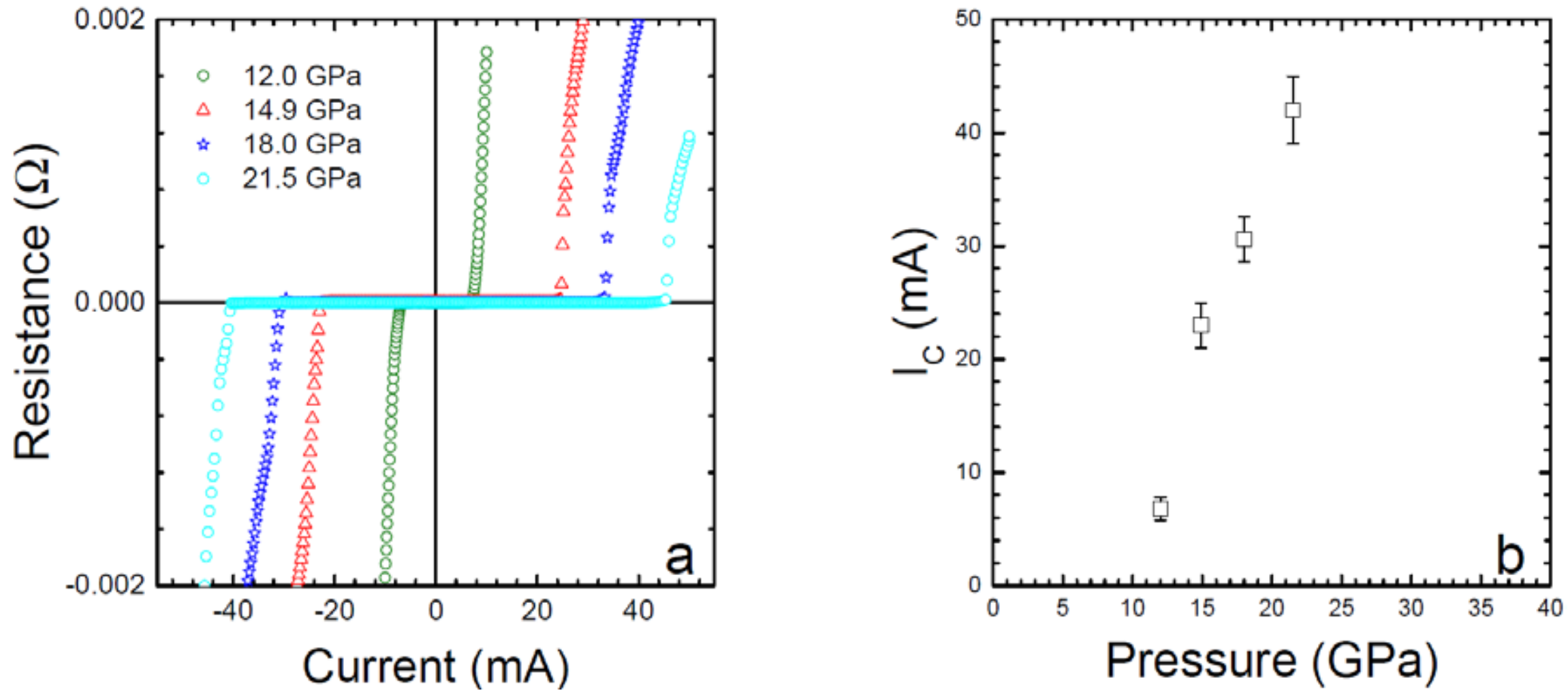

**Supplementary Figure S3. Superconducting critical current at 2K as function of pressure of a-GST. a,** Resistance as a function of current at 2K at different applied pressure. **b**, Critical current as a function of pressure.

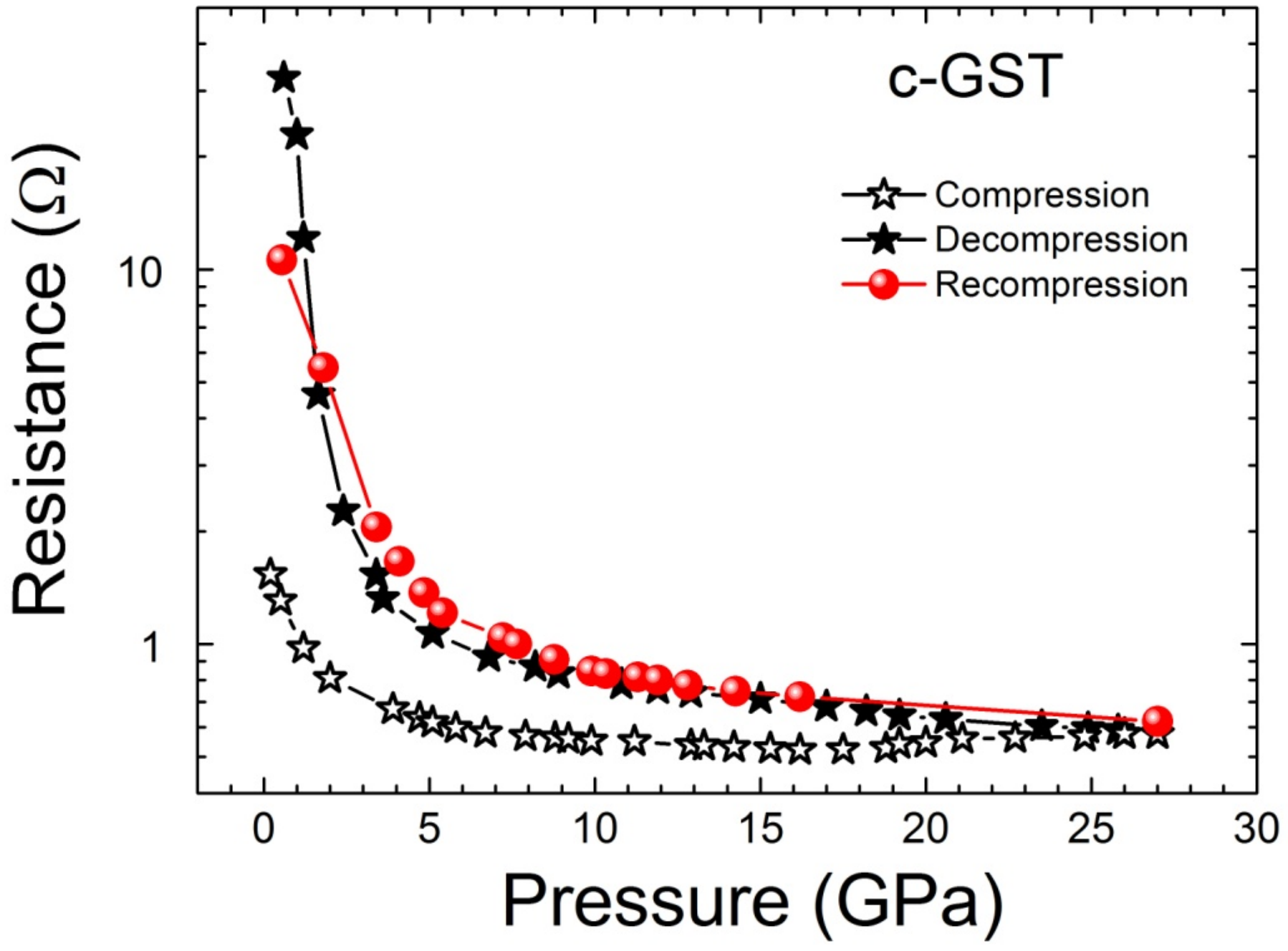

**Supplementary Figure S4. Room temperature resistance under pressure for c-GST during compression, decompression and re-compression cycle.**

**Supplementary Methods**

One micron thick *a*-GST films were deposited from a stoichiometric hexagonal $GeSb_2Te_4$ (*h*-GST) target using magnetron sputtering in Argon plasma (base pressure ~$10^{-7}$Torr) at Ar pressure of $5\times10^{-3}$ Torr. *c*-GST in its hexagonal phase (*h*-GST) was prepared in the form of powder from the same material as the target used for the sputtering of *a*-GST.

Pressure is exerted using miniature diamond anvil cells (DACs) [i] with diamond anvil culets of 300 μm. A preindented stainless-steel or rhenium gasket was drilled and then filled and covered with a powder layer of 75% $Al_2O_3$ and 25% NaCl for electrical insulation. The *a*-GST films were mechanically removed and placed in the form of powder onto the culets, with 4 triangular Pt contacts allowing the electronic transport measurements at elevated pressures up to 36 GPa, as depicted in figure 1 c. The *h*-GST was placed into a similar DAC as described above. Ruby was used as a pressure gauge.

For the XRD measurements we used a DAC with a Boehler-Almax anvil, a stainless steel gasket, and a 4:1 methanol-ethanol hydrostatic medium. For *a*-GST XRD data was taken using a Bruker Discovery 8 system with a Microfocus source of molybdenum (K-alpha= 0.7106 Å) with a beam collimated to 110-micron diameter. Data was collected typically for 20 min. XRD measurement of *h*-GST powder was performed on ID27 beamline at the European Synchrotron Radiation Facility (ESRF), Grenoble, France (λ = 0.3738 Å) in angle-dispersive mode with patterns collected using a Perkin Elmer flat panel detector.

Electrical transport measurements were performed using a physical property measurement system (PPMS Quantum Design, USA).

---

[i] Sterer, E., Pasternak, M. P. & Taylor, R. D. *Rev. Sci. Instrum.* **61,** 1117 (1990).